\documentclass[prl,twocolumn]{revtex4}
\usepackage{graphicx}
\usepackage{amsmath}
\usepackage{amssymb}
\usepackage{color}
\usepackage[normalem]{ulem}
\usepackage{epsfig}

\begin{document}
\title{Spatial neutral dynamics}
\author{Matan Danino\footnote{These authors contributed equally to this work.}, Yahav Shem-Tov$^*$ and Nadav M. Shnerb }
\affiliation{Department of Physics, Bar-Ilan University, Ramat-Gan
IL52900, Israel}

\begin{abstract}
\noindent Neutral models, in which individual agents with equal fitness undergo a birth-death-mutation process, are very popular in population genetics and community ecology. Usually these models are applied to populations and communities with spatial structure, but the analytic results presented so far are limited to well-mixed or mainland-island scenarios.
Here we present a new technique, based on interface dynamics analysis, and apply it to the neutral dynamics in one, two and three spatial dimensions. New results are derived for the correlation length and for the main characteristics of the community, like total biodiversity and the species abundance distribution above the correlation length. Our results are supported by extensive numerical simulations, and provide qualitative and quantitative insights that allow for a rigorous comparison between model predictions and empirical data.
\end{abstract}

\maketitle

Neutral dynamics, and the neutral models used to describe it, are one of the main conceptual frameworks in population biology and ecology \cite{kimura1985neutral,hubbell2001unified,azaele2015statistical}. A neutral community is a collection of different populations, such as different species (in ecological models) or different groups of individuals with identical  genetic sequence (haplotypes, for example, in population genetics). All individuals undergo a stochastic birth-death process, where in most of the interesting scenarios the overall size of the community, $J$, remains fixed or almost fixed. An offspring of an individual will be a member of its parent group with probability $1-\nu$, and with  probability $\nu$ it mutates or speciates, becoming the originator of a new taxon.  A  neutral process  does not include selection: all populations are demographically equivalent, having the same rates of birth, death and mutations, and the only driver of  population abundance variations is the stochastic birth-death process (demographic noise).

 A neutral dynamics is relevant, of course, to any inherited feature that does not affect the phenotype of an individual, such as a polymorphism in the non-coding part of the DNA or silent mutations, but many believe that its scope is much wider. In particular, the neutral theory of molecular evolution \cite{kimura1985neutral} and the neutral theory of biodiversity \cite{hubbell2001unified} both suggest that even the phenotypic diversity observed in natural communities reflects an underlying neutral or almost-neutral process while the effect of selection is absent or very weak. Both theories have revolutionized the fields of population genetics and community dynamics, correspondingly, and despite bitter disputes, their  influence is overwhelming.

For a well-mixed ($0$d) community the mathematical analysis of the neutral model is well-established, with the theory of coalescence dynamics \cite{wakeley2009coalescent} and Ewens's sampling formula \cite{ewens1972sampling} at its core. However, the species abundance distribution predicted by this model, the Fisher log-series, fails to fit the observed statistics of trees in a tropical forest. To overcome this difficulty, Stephen Hubbell suggested a simple spatial generalization of the neutral model, where a well mixed community on the mainland (a "metacommunity") is connected to a relatively small island by migration and immigrant statistics is given by Ewens's sampling formula \cite{hubbell2001unified,maritan1}. The abundance of a species on the island reflects the balance between its mainland abundance (assumed to be fixed, as variations on the mainland are much slower) and local stochasticity.  The  resulting island statistics depend on two parameters only, the combination $\theta = \nu J_m$ ($J_m$ is the mainland abundance) and $m$, the migration rate. The success of this two-parameter model in describing local communities,  and its mathematical simplicity that allows for an exact solution in terms of zero-sum multinomials \cite{maritan1}, were the key ingredients that contributed to the success of Hubbell's neutral theory \cite{rosindell2011unified,azaele2015statistical}.

Still, this mainland-island model is only an approximation. The tropical forest plots used to validate it are not "islands" per se, instead they are arbitrary segments of very large forests on which a census takes place. Even the plot known as "Barro-Colorado Island" is a $500 \times 1000$m rectangle where the island area is  $15.6 km^2$. In practice  there is no natural distinction between the local population and its surroundings and local dispersal ensures correlations between the two, correlations that have no analog in Hubbell's mainland-island model. Consequently, one would like to have a solution, or at least a set of intuitive arguments, for the generic problem of spatially explicit neutral dynamics \cite{ter2010neutral}. Several attempts have been made in this direction, both in the context of community ecology~\cite{zillio2005spatial,azaele2015towards,azaele2015statistical,o2010field,de2009global,rosindell2007species} and in the context of population genetics~\cite{wilkins2002coalescent}.

The aim of this letter is to present a novel analysis, based on interface dynamics, of the spatial neutral model. Armed with this tool we can present expressions for the correlation length, species abundance distribution and species richness, and these expressions are shown to fit very nicely the results of extensive numerical simulations.

Technically speaking, the neutral dynamics is a "technicolor" version of the well known voter model~\cite{liggett2013stochastic}. In the original voter model any individual has one of two colors, or opinions, and in an elementary timestep an agent is chosen at random to change its color, accepting instead the color of one of its randomly chosen neighbors. Such a game ends up, inevitably, with a fixation of the population by one color. A neutral game proceeds according to the same rules, with the exception that the agent accepts its neighbor's opinion with probability $1-\nu$ and, with probability $\nu$, it becomes the originator of a new color (note that, unlike the two allele model considered by~\cite{korolev2010genetic}, in the infinite allele case considered here recurrent mutations are not allowed and a brand  new species appears in every mutation).

\begin{figure}
\centering{
\includegraphics[width=8cm]{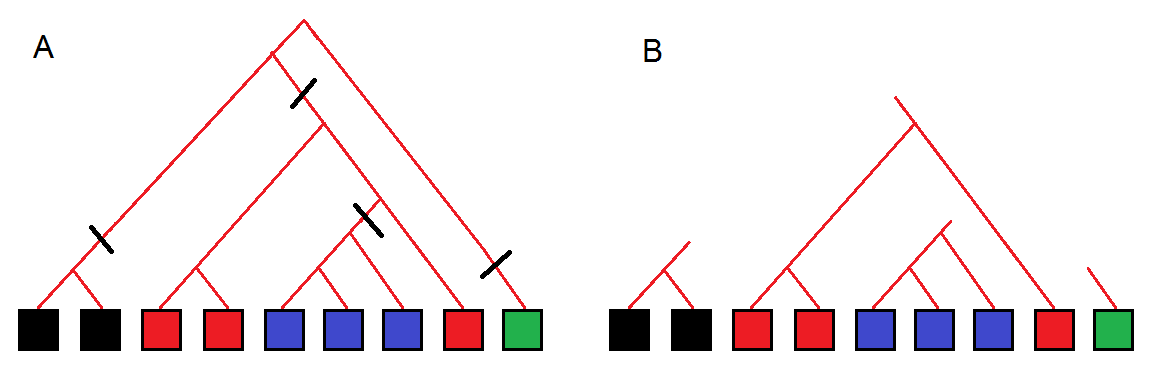}}
\caption{Two cartoons showing a possible genealogy and its corresponding neutral dynamics. In (A) the full genealogy of a $1d$ coalescence process with 9 individuals is presented. Every mutation  (represented by thick dark lines cutting the line) generates a new species, and conspecific individuals at present have the same color.  Panel (B)  shows the same genealogy when simulated only until the last relevant mutation: coalescence history and  mutations in the "missing" part of the tree are irrelevant.} \label{fig1}
\end{figure}

Like the traditional voter model, the neutral dynamics may be analyzed using a "backward in time" (coalescence) approach, becoming a coalescence random walk ($A+A \to A$) process \cite{ben2000diffusion}. In its nearest-neighbor spatial version every individual selects its parent from one of its  neighbors, and coalescence occurs when two agents choose the same parent. The resulting  genealogic tree (in $1d$) is illustrated in Fig \ref{fig1}a, where lines representing ancestral relationships  merge until the dynamics reaches the most recent common ancestor. Mutation/speciation events are represented by short thick lines that cut the lines, and all the leaves connected to a certain mutant by lines without mutations carry the same color, i.e, they belong to the same species. This property  facilitates the simulation of a neutral dynamics~\cite{rosindell2007species}: instead of simulating the coalescence random walkers until the most recent common ancestor and then introducing random mutations with a chance proportional to  the length of a line (number of birth), we have simulated coalescing and dying ($A \stackrel{\nu}{\longrightarrow} \emptyset$) walkers,  as depicted in figure \ref{fig1}b.
 Monitoring all descendent of an individual we were able to easily identify the colors of agents at $t=0$ (leaves) and to simulate a large system until it reaches the most recent \emph{relevant} mutation (i.e., mutation that yields a currently existing species), avoiding the diverging timescales associated with the last coalescence events when only a few ancestors survive.

 While implementing a backward in time approach for the numerics, our analytic arguments are based on the forward in time evolution of the system and focus on the interface area. To begin, let us consider  the neutral dynamics in $1d$ (a model considered by geneticist, see \cite{wilkins2002coalescent}). Looking at a species represented by $x$ individuals (e.g., for the red species in Fig. \ref{fig1}, $x=3$,  for the black $x=2$), one realizes that its dynamics is governed by two processes: weak losses at a rate $\nu x$ per generation due to mutations, and an unbiased diffusion in abundance space associated with the birth-death dynamics.  Clearly, the strength of diffusion for $x$ is proportional to $I$, the  interface between single color segments.  For example, the red species in Fig. \ref{fig1} has four interfaces, while for the blue one $I=2$.  Assuming a narrow distribution of the number of interfaces around an average $I(x)$ (our simulations suggest a Poisson distribution), one may write a Fokker-Planck equation for the single species abundance dynamics,
\begin{equation} \label{eq1}
\frac{\partial P(x,t)}{\partial t} = \frac{\partial^2 [I(x)P(x,t)]}{\partial x^2} + \nu \frac{\partial [xP(x,t)]}{\partial x}
\end{equation}
where $P(x,t)$ is the probability of a certain species to be represented by $x$ individuals at $t$ ($t$ is measured in generations). Once $I(x)$ is known, the equilibrium species abundance distribution (SAD) is given by,
\begin{equation} \label{eq1a}
P_{eq}(x) = \frac{e^{ -\nu \int dx \frac{x}{I(x)}}}{I(x)}.
\end{equation}
This formula is valid in any dimension, but $I(x)$ depends on dimensionality. To suggest an expression for $I(x)$, we introduce here a few arguments. The $1d$ case is discussed first, but some of the insights will be used below for higher dimensions.

First, the correlation length $\xi$ is defined via the chance of two individuals, at a distance $\ell$ apart, to have the same color. Having the backward picture in mind, this is equivalent to the chance that two random walkers,  starting at a distance $\ell$ from each other, will coalesce before mutation occurs, i.e., within a time shorter than $1/(2\nu)$. The  theory of first passage time~\cite{redner2001guide} suggests that, for a nearest neighbors $1d$ dynamics, $\xi \sim 1/\sqrt{\nu}$. Accordingly, one should expect that species with $x>\xi$ will be rare (i.e., that $P_{eq}^{1d}(x)$ drops substantially above $\xi$) and that $I(x)$  scales linearly with $x$ for $x \gg \xi$, as the density of gaps becomes uncorrelated.

A second argument has to do with  the overall species richness (SR). For a system of coalescing random walkers in one dimension the density of agents, $n(s)$  is known to fall like $n(s) = n(0)/\sqrt{s}$~\cite{ben2000diffusion}. Here $s$ represents time, measured in generations, and we use $s$ instead of $t$ since in our case the coalescence picture is relevant when time is measured backward, starting with  $n(0) = J$, the overall size of the community. The SR of a sample is given by integration over $s$, where in each generation one counts the number of agents that survived the coalescence-death process, $J e^{-\nu s}/s$, and multiplies it by the chance for mutation. The answer is $\nu$ times the volume of the "truncated" tree shown in Figure \ref{fig1}b,
\begin{equation} \label{eq2}
SR(J,\nu) \approx \nu J \int_1^\infty \frac{e^{-\nu s}}{\sqrt{s}}\ ds  = B \sqrt{\nu} J
\end{equation}
where $B$ is a constant of order unity.

On the other hand, $SR(J,\nu)$ is related to the SAD, $P_{eq}(x)$. The linear equation (\ref{eq1}) yields a non-normalized SAD where normalization is determined by the condition  $\int x P_{eq}(x) dx = J$.  Accordingly~\cite{zillio2005spatial},
\begin{equation} \label{eq3}
SR(J,\nu)   = B \sqrt{\nu} J =  J \frac{\int_1^\infty P_{eq}(x) \ dx}{ \int_1^\infty  x P_{eq}(x) \ dx}.
\end{equation}
 A simple scaling argument shows that (if $P_{eq}$ is non-singular at zero)  $P_{eq}$ is a function of $\sqrt{\nu} x$. Combining this  with Eq. (\ref{eq1a}) and with the first argument, we suggest $I_{1d}(x) \sim 2 + 2 B \sqrt{\nu} x {\cal F} (\sqrt{\nu} x)$ where ${\cal F}(z \gg 1) \to 1$.

\begin{figure}
\centering{
\includegraphics[width=7cm]{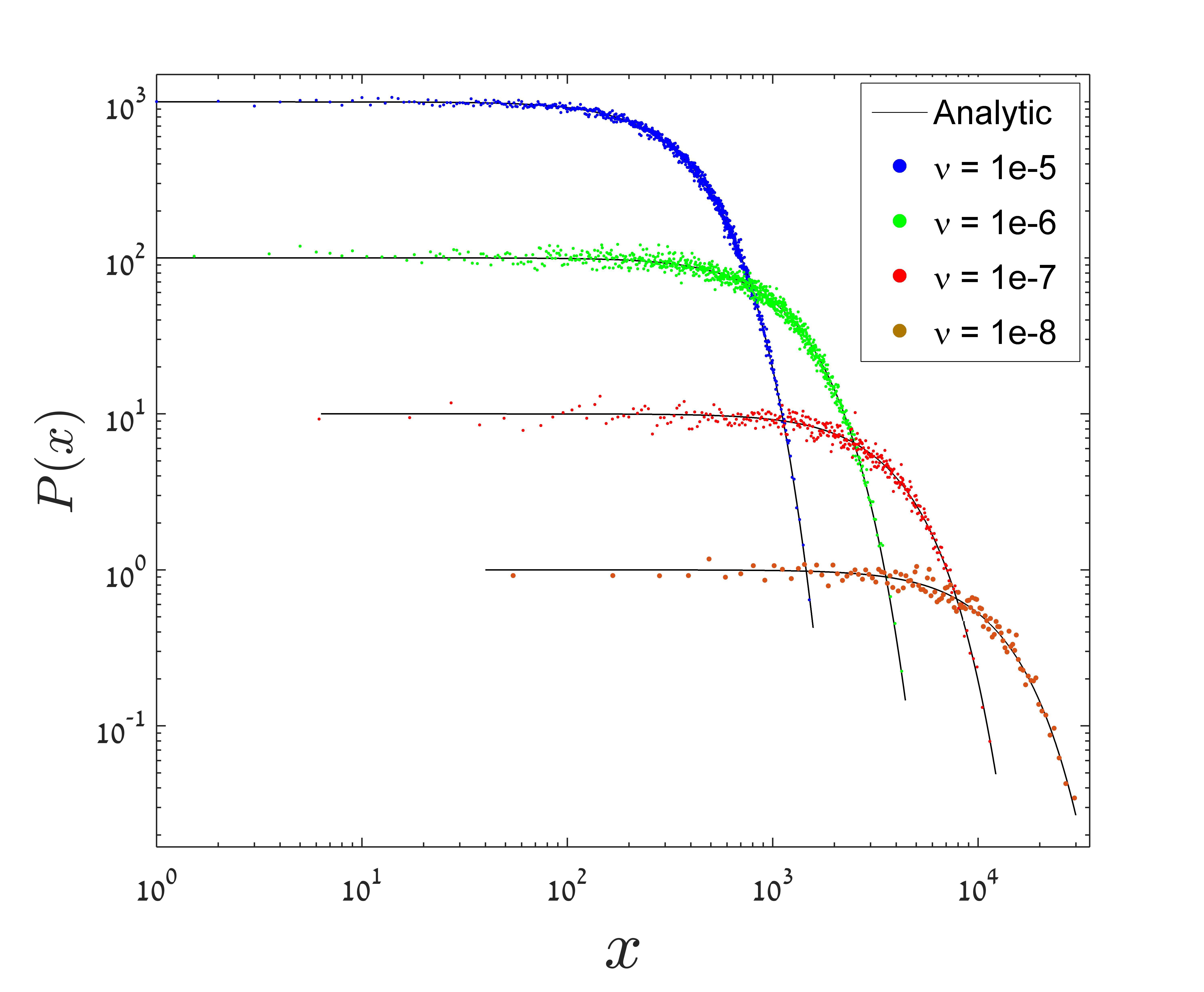} \\
\includegraphics[width=7cm]{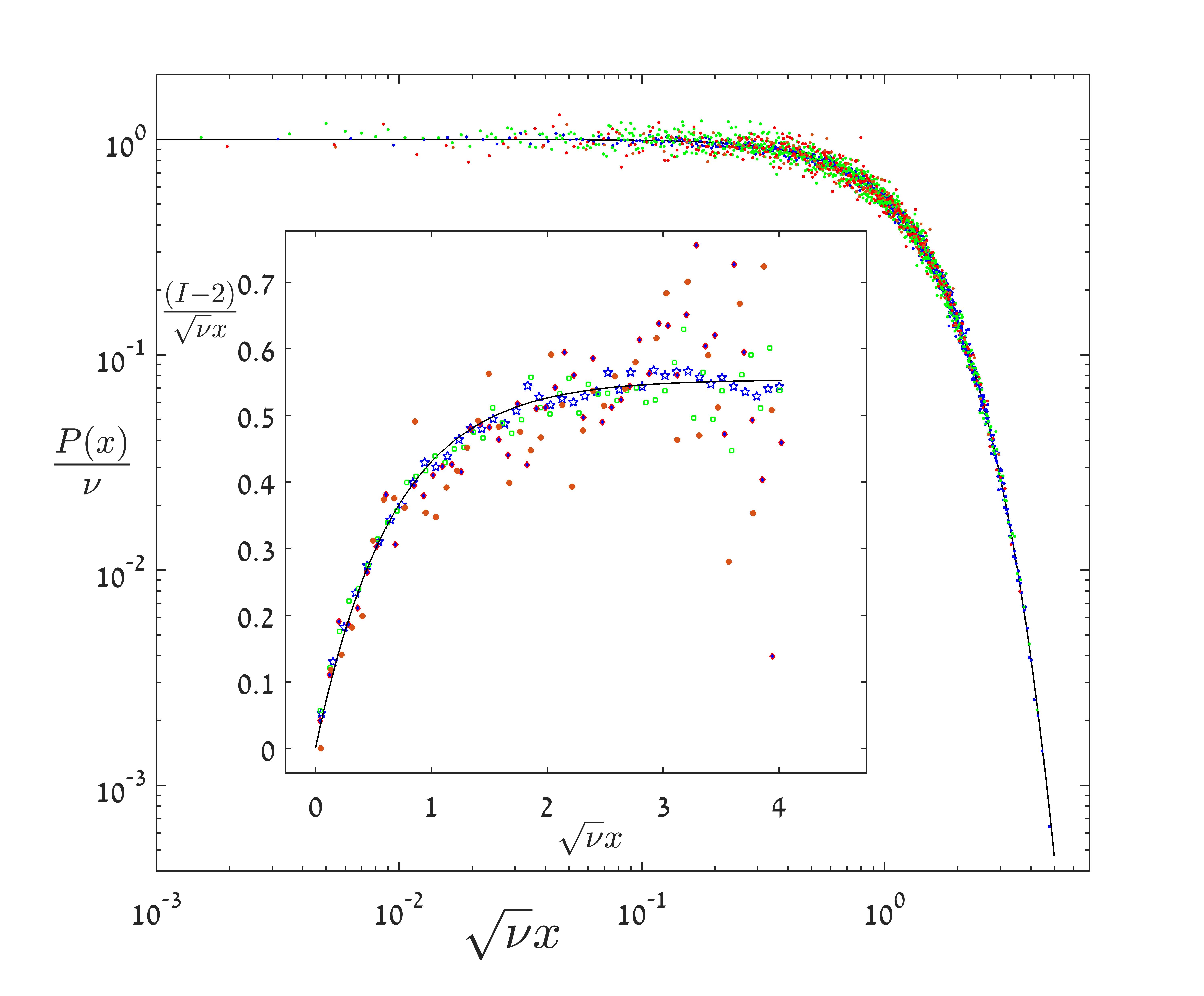}}
\caption{Species abundance distribution, $P(x)$ vs. $x$, for a $1d$ model with $J = 10^8$ and different values of $\nu$, ranging from $\nu = 10^{-5}$ to $10^{-8}$ (upper panel). The agreement with Eq. \ref{eq4}, with $A = 1.5$ and $B = 0.27$, is evident.  In the lower panel $P(x)/\nu$ is plotted against $\sqrt{\nu}x$ for all datasets, and the perfect data collapse indicates that  $P(x)$ is  a function of $\nu {\cal G}(\sqrt{\nu} x)$, where ${\cal G}$ is a universal scaling function. The inset shows how $(I_{1d}(x)-2)/\sqrt{\nu}x$ scales with $\sqrt{\nu}x$, depicting the crossover from linear to saturation and the fit to the (full line) $\left( 1-\exp(-A \sqrt{\nu} x) \right)$ (with the same value of $A$ for all $\nu$s). Datapoints for $x>\xi$ are rare, so the plot for these values is more noisy.    } \label{fig2}
\end{figure}

 Species with $x \ll \xi$ are quite compact. When a mutation occurs, it yields two interfaces that behave like two random walkers starting at $\ell = 1$. If they survive until $t$ the corresponding species have abundance $x \sim \sqrt{t}$,
    so their genealogy is contained, more or less, within a triangle of base size $x$  and "height" $x^2$.  The chance of a mutation  inside this triangle to occur $s$ generations before present is $\nu \sqrt{x^2 -s}$,   and it survives (hence adding two interfaces) with probability $1/\sqrt{s}$. Accordingly,  the expected number of interfaces is given by  $\nu \int_0^{x^2} ds \sqrt{(x^2-s)/s} \sim \nu x^2$ and  we thus expect ${\cal F}(x \ll \xi) \to \sqrt{\nu} x$. Assuming further an exponential convergence to the $x>>\xi$ asymptotic value, one may guess a capacitor charging form $$I_{1d}(x) = 2+ 2B \sqrt{\nu} x \left( 1-\exp(-A \sqrt{\nu} x) \right)$$, with $A$ another constant. Plugging this into Eq.~(\ref{eq1a}) one obtains,
\begin{equation} \label{eq4}
P_{eq}^{1d}(x) = C \frac{e^{-  \int\limits^{\sqrt{\nu}x} \frac{y\ dy}{2 \left[1+By(1-exp(-Ay)) \right]}}}{2 \left[1+B\sqrt{\nu}x(1-exp(-A\sqrt{\nu}x))\right]}.
\end{equation}
When $x \ll \xi$ the number of interfaces is almost fixed at two and the decay is Gaussian, but for large $x$ the decay switches to an exponential form. Figure \ref{fig2} shows $P_{eq}^{1d}(x)$ for certain values of $\nu$, emphasizing the excellent fit and the data collapse when $P_{eq}^{1d}/\nu$ is plotted against $\sqrt{\nu}x$, where the constant was found to be $C = \nu J$.

Establishing this intuitive framework by studying the $1d$ case, let us consider now the (much more important) $2d$ neutral model. Two is the critical dimension of the coalescing random walk problem \cite{ben2000diffusion,krapivsky1992kinetics} and of the first passage time in general \cite{redner2001guide}, with logarithmic corrections to the mean field results, so one may expect that its analysis will be more difficult. This is probably true if the problem has to be solved exactly. However, for the  analysis considered here the $2d$ model appears to be easier than its $1d$ counterpart.

Under a simple voter-model dynamics without mutations, the chance of the lineage of an individual to survive after $t$ generations goes like $ln(t)/t$ (as opposed to $1/t$ above $2d$) \cite{krapivsky1992kinetics}.  Accordingly, to keep the population fixed  the average number of offspring of a surviving individual  after $t$ generations has to be $t/ln(t)$. Therefore, up to logarithmic corrections, the age of a species with abundance $x$ is $t(x) \sim x\ln(x)$

\begin{figure}
\centering{
\includegraphics[width=7cm]{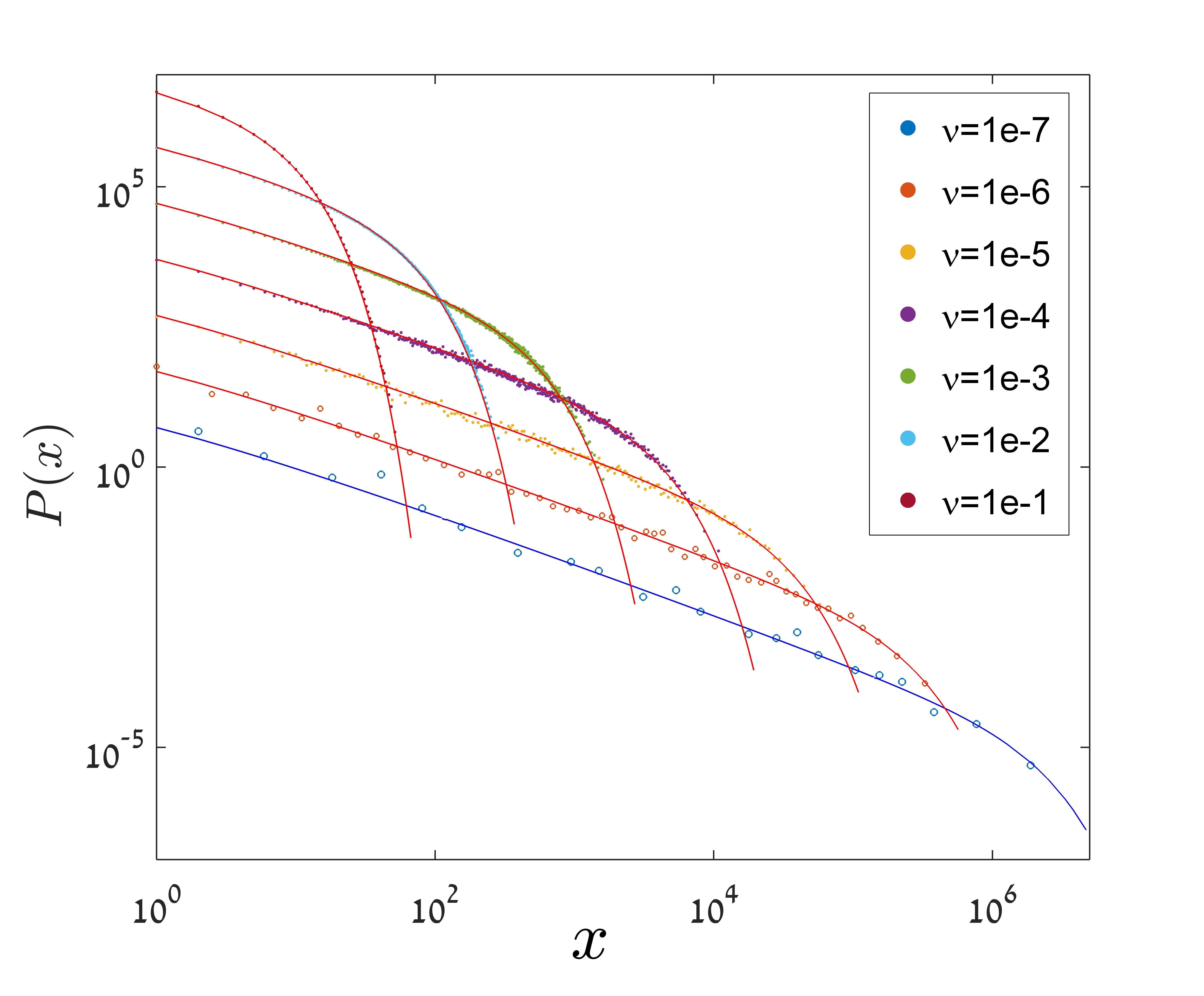} \\
\includegraphics[width=7cm]{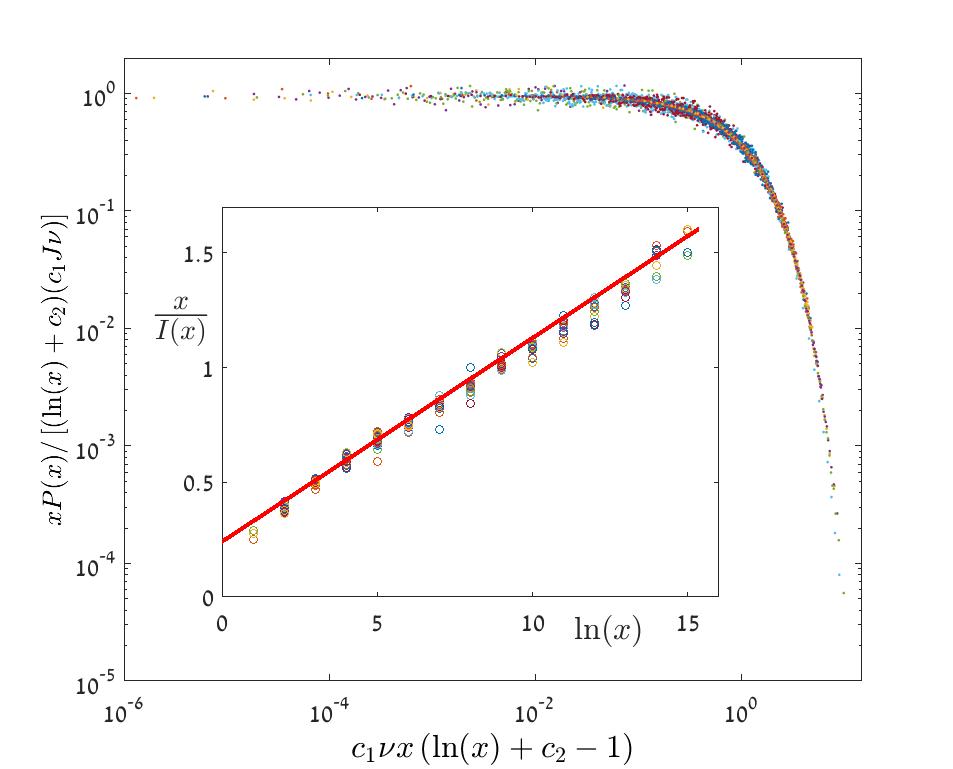}}

\caption{The upper panel shows the species abundance distribution, $P(x)$ as a function of $x$, for $\nu$ values between $10^{-1}$ to $10^{-7}$,  together with fits to Eq. (\ref{eq5}), with $c=0.36$. In the lower panel the data collapse obtained when $xP_{eq}/[1+c\ \ln(x)]$, for all these curves, is plotted against $\nu x[1+ c (\ln(x)-1)]$.  The inset shows  $x/I(x)$ vs. $\ln(x)$ for $31$ runs from $\nu = 10^{-8}$ to $\nu = 10^{-1}$ (circles) and the full straight line indicates that $I(x) =x/(1+c\ \ln(x))$, with the same $\nu$ independent value of $c$. Datasets were collected from simulations of the backward in time process for systems of size  $7000 \times 7000$ ($J = 4.9 \cdot 10^7$)  with a nearest neighbor dispersal kernel.} \label{fig3}
\end{figure}

Now, the neutral dynamics without mutation satisfies Eq. (\ref{eq1}) with $\nu=0$. A simple scaling argument shows that to have $t \sim x \ln(x)$,  $I_{2d}(x) = x/(1+c\ \ln(x))$, where $c$ is constant related to the amplitude of the kernel. Plugging this expression into Eq.(\ref{eq1a}) the SAD is found to be,
\begin{equation} \label{eq5}
P_{eq}^{2d} = A \frac{1+ c\ \ln(x)}{x} \exp\left(- \nu x[1+ c (\ln(x)-1)]\right).
\end{equation}
Figure \ref{fig3} shows the excellent fits and data collapse this formula yields when $\nu$ varies over six decades, with  $A = \nu J$. \color{black} The species richness above $\xi$ is determined by Eq. (\ref{eq2}); while the integration itself yields a complex expression, we find that

\begin{equation}\label{eq5a}
SR^{2d}(J,\nu) \sim J \nu [\ln(\nu)-1]^2,
\end{equation}

yields a decent approximation (maximum error of $15\%$ at $\nu = 0.01$, converging to the correct result as $\nu$ becomes smaller). This expression also converges to $J$ when $\nu \to 1$.

Above the critical dimension, $d>2$, the mean-field expressions of Galton-Watson theory describe accurately the dynamics at long times, meaning that the chance of the lineage of a new mutant to survive $t$ generations is $1/t$ and a surviving mutant has $t$ offspring. Accordingly, from Eq. \ref{eq1} with $\nu=0$, $I(x)$ must scale linearly with $x$. This leads immediately to the celebrated Fisher log-series statistics,
 \begin{equation} \label{eq6}
P_{eq}^{d>2} =\frac{A e^{-\nu x}}{x}.
\end{equation}
 This Fisher log-series has been implemented in the neutral models \cite{hubbell2001unified,maritan1} as the SAD on the mainland. In  the relevant parameter regime the 2d SAD (\ref{eq5}) and the mean field expression (\ref{eq6}) differ strongly, both for frequent species (where the exponential decay is replaced by a factorial decay) and in the tail, where logarithmic corrections are important.

The results presented here disagree with the scaling analysis suggested in \cite{zillio2005spatial} (this scaling was already criticized in \cite{pigolotti2009speciation}, where it failed to fit numerical results) . As one realizes from the backward in time exposition of the problem, the oldest species were originated about $1/\nu$ generations before present, and since every single lineage preforms an unbiased random walk, the largest distance between two conspecific individuals, which sets the correlation length, is of order $1/\sqrt{\nu}$, up to logarithmic corrections in $2d$.   By the same token, the field theoretical analysis presented in \cite{o2010field}, and in particular the expression suggested for the species area curve (Eq. (10) of \cite{o2010field}) are in contrast with our simulation results and with Eq. (\ref{eq5a}), as these authors predict a purely linear dependence of the SR on $\nu$, and a $\nu$ independent correlation length.

 Our results, and in particular the universal characteristics of the community  such as the functional dependence of species age on its abundance and the tail of the SAD, appear to be relevant for the new  generation of large scale spatial surveys, like those presented recently for tropical forests \cite{ter2013hyperdominance,slik2015estimate}. The  data analysis in these works  depends strongly on the assumption that the SAD is  Fisher log-series; reinterpretation of these results in view of the spatially explicit model and its SADs presented here could be an enlightening exercise.

{\bf Acknowledgments} We acknowledge the support of the Israel
Science Foundation, grant no. $1427/15$.

\bibliography{references}

\end{document}